\documentclass[journal=jacsat,manuscript=article]{achemso}

\usepackage[version=3]{mhchem} 

\author{Axel J. M. Deenen}
\author{Dirk Grundler}
\affiliation{Laboratory of Nanoscale Magnetic Materials and Magnonics, Institute of Materials (IMX), School of Engineering, École Polytechnique Fédérale de Lausanne (EPFL), Lausanne, 1015, Vaud, Switzerland}
\altaffiliation{Institute of Electrical and Micro Engineering, School of Engineering, École Polytechnique Fédérale de Lausanne (EPFL), Lausanne, 1015, Vaud, Switzerland}
\email{Dirk.Grundler@epfl.ch}

\title[An \textsf{achemso} demo]
  {Periodic Phase Slips and Frequency Comb Generation at Tunable Microwave Frequencies in Superconducting Diabolo Structures}

\begin{document}

\begin{abstract}
Superconductors are characterized by macroscopic phase coherence and have enabled cryogenic electronics and quantum technologies. Recent advances in 3D nanofabrication now offer possibilities for tuning functional properties relevant for on-chip 3D integration of superconductors. However, non-equilibrium phenomena in 3D nanostructures exposed to transport currents remain largeley unexplored. Here, we employ numerical simulations to investigate phase slips ---discrete $2\pi$ jumps in the phase of the superconducting order parameter--- in a tubular Nb superconductor with a central constriction, which is subjected to both direct current (DC) and alternating current (AC) transport currents. We find that under DC drive, the system stabilizes periodic phase slips, resulting in GHz voltage oscillations. Introducing an additional AC frequency modulation generates microwave frequency combs which depend characteristically on the interaction between moving vortices and phase slips. Our  findings open avenues for developing on-chip frequency comb generators in 3D cryoelectronics.

\end{abstract}

\section{Introduction}
\begin{figure}[h]%
\centering
\includegraphics[width=0.95\textwidth]{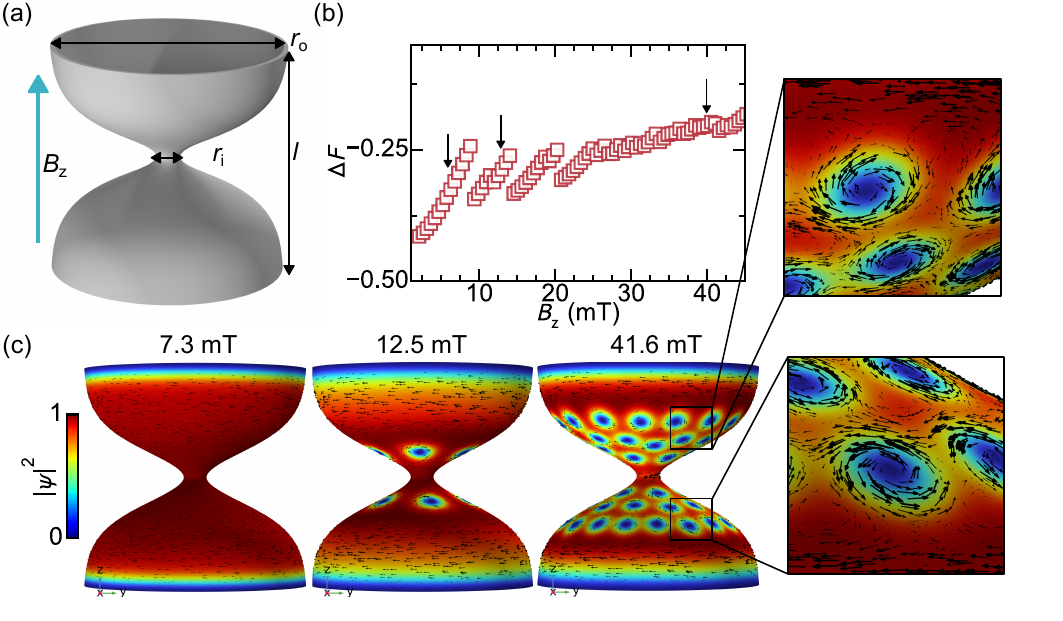}
\caption{States of the diabolo under constant applied axial field. (a) Geometry of the diabolo highlighting the inner radius $r_i$, the outer radius $r_o$, the length $l$ and the applied field $B_z$. (b) The computed free energy difference $\Delta F$ between the superconducting state and the normal state as a function of field. (c) The spatial distribution of the order parameter magnitude at several fields indicated by arrows in (b).
}
\label{fgr:fig1}
\end{figure}

Superconductors are at the forefront of nanoscience and quantum technology, underpinning innovations in qubits, ultra-sensitive sensors, and parametric amplifiers. Central to many of these technologies is the Josephson junction, typically formed by two superconductors (S) separated by a thin insulating barrier (I). Recently, generation of frequency combs using Josephson junctions and a superconducting resonator was demonstrated \cite{wang_integrated_2024}.\\
\indent Despite their success, SIS junctions are inherently limited by their planar geometry and fabrication requirements, which necessitate distinct materials for the superconducting and insulating layers. An alternative approach involves the use of weak links\cite{likharev_superconducting_1979,lindelof_superconducting_1981}. A typical example consists of two superconducting banks connected by a superconducting constriction, often refered to as a micro- or nano-bridge\cite{klapwijk_regimes_1977}. Typically, when the width and length of these constrictions are smaller than the coherence length ($\xi$), Josephson effects can be observed. \\
\indent Additionally, a current-driven superconductor with a weak link can exhibit a resistive state characterized by periodic suppression of the order parameter magnitude and $2\pi$ phase slips\cite{ivlev_electric_1984, kimmel_phase_2017}. Phase slips manifest as phase-slip centers\cite{skocpol_phase-slip_1974} in quasi-one-dimensional systems or as phase-slip lines\cite{andronov_kinematic_1993} in wider films and induce quantum interference effects similar to Josephson junctions\cite{sivakov_josephson_2003}. They have been proposed for use in flux qubits\cite{mooij_phase-slip_2005} and optimizing transmon qubits\cite{liu_performance_2024}. Recent advances in nanofabrication have enabled the realization of 3D superconducting architectures with nanoscale precision\cite{makarov_new_2022, fomin_perspective_2022, cordoba_topological_2024}, making controllable constriction-based phase slips especially promising for high-density 3D sensing and computing architectures\cite{collins_superconducting_2023}.\\
\indent However, while phase slip phenomena in one dimension are relatively well understood\cite{altomare_onedimensional_2013, yerin_dynamics_2013}, analytical descriptions in two and three dimensions remain challenging due to the complexity of the governing equations. In particular, the interplay between topological defects, such as vortices and phase slips\cite{bogush_microwave_2024, fomin_perspective_2022}, in 3D nanostructures is not fully understood.\\
\indent In this work, we conduct full 3D time-dependent Ginzburg-Landau simulations to investigate phase slip phenomena in Nb diabolo structures --- hollow tubes featuring a central constriction (Figure \ref{fgr:fig1}a) --- under both DC and combined DC+AC current drives. We demonstrate that the intrinsic frequency of phase slip oscillations, given by the Josephson relation \cite{watts-tobin_nonequilibrium_1981}, can be harnessed to generate microwave frequency combs, potentially advancing high-frequency quantum applications. Furthermore, we demonstrate the coexistence of phase slips with vortices and show that at sufficiently high currents, vortices are absorbed by the phase slip region, offering insights into controlling topological defects in superconducting systems.
\\

\section{Results and discussion}
We have numerically simulated several tubular structures of different geometrical parameters and report here on a superconducting Nb diabolo structure (Figure \ref{fgr:fig1}a). It consists of an Nb shell with a thickness of $d = 27$ nm and length $l = 1.64~\mu$m. The radius varies from $r_{\rm o} = 890$ nm at the ends to $r_{\rm i} = 58$ nm at the central constriction. Throughout this work, we consider a fixed temperature $T=0.95T_{\rm c}$, with $T_{\rm c}$ the critical temperature. The penetration depth and coherence length are taken as $\lambda = 273$ nm and $\xi = 58$ nm, respectively, giving a Ginzburg-Landau parameter $\kappa=\lambda/\xi=4.7$\cite{rezaev_topological_2020}. The normal conductivity $\sigma=16~\rm{(\mu\Omega)^{-1}}$. Since $d \ll \lambda$, self-field effects are neglected. To study the dynamic response, we numerically solve the time-dependent Ginzburg–Landau (TDGL) equations using a 3D finite-element formulation\cite{du_finite_1994} implemented in COMSOL Multiphysics\cite{noauthor_comsol_nodate, oripov_time-dependent_2020}. We adopt the Coulomb gauge\cite{gao_finite_2023}, imposing $\nabla\cdot\mathbf{A}=0$ on the vector potential $\mathbf{A}$.  

Within the Ginzburg–Landau framework, the dimensionless free energy difference between the normal and superconducting states, expressed in terms of the complex order parameter $\psi$, is given by:
\begin{equation}
    \Delta F = -|\psi|^2 + \frac{1}{2}|\psi|^4 + \left| \left( \frac{1}{\kappa}\nabla - i\mathbf{A} \right) \psi \right|^2 + |\nabla \times \mathbf{A} - \mathbf{H}|^2.
    \label{eq:GL}
\end{equation}
Here, lengths are normalized to $\lambda$, and time to $\xi^2/D$, with $D$ being the diffusion coefficient. The last term in Eq.~\ref{eq:GL} is negligible in the limit of small self-fields. The evolution of $\psi$ is described by the TDGL equation\cite{schmid_time_1966, du_finite_1994}:
\begin{equation}
    \left( \frac{\partial}{\partial t} + i\kappa\phi \right)\psi = \left( \frac{1}{\kappa}\nabla - i\mathbf{A} \right)^2\psi + (1 - |\psi|^2)\psi,
\end{equation}
supplemented by the Poisson equation for the scalar potential $\phi$ to ensure current conservation:        
\begin{equation}
        \sigma\nabla^2\phi=-\nabla\cdot\left(\frac{i}{2\kappa}\left(\psi^*\nabla\psi-\psi\nabla\psi^*\right)+|\psi|^2\mathbf{A}\right),
\end{equation}
where $\sigma$ is the normal conductivity.  
Boundary conditions are applied as follows:  
\begin{itemize}
    \item At the superconductor/vacuum (SV) interface: $\nabla\psi\cdot\hat{n} = 0$, $\nabla\phi\cdot\hat{n} = 0$, and $\mathbf{A}\cdot\hat{n} = 0$.  
    \item At the superconductor/metal (SN) interface: $\psi = 0$ and $\nabla\phi\cdot\hat{n} = -j_{\rm tr}/\sigma$.  
\end{itemize}
With $j_{\rm tr}$ the applied transport current. Additionally, to ensure uniqueness, we impose the condition:
\begin{equation}
\int_\Omega \phi \, d\Omega = 0,
\end{equation}
where $\Omega$ denotes the simulated domain.  

An external magnetic field $B_{\rm z}$ is applied along the diabolo’s central axis ($z$-axis). The vector potential $\mathbf{A}$ satisfies the gauge condition and boundary conditions and is expressed as:
\begin{equation}
\mathbf{A} = \mathbf{A}_{\rm ext} + \nabla\chi,
\end{equation}
where $\mathbf{A}_{\rm ext} = \frac{1}{2}B_z x\hat{y} - \frac{1}{2}B_z y\hat{x}$ corresponds to the symmetric gauge, and $\chi$ is a gauge function solving the Laplace equation:
\begin{equation}
\Delta\chi = 0,
\end{equation}
subject to the boundary condition:
\begin{equation}
\nabla\chi\cdot\hat{n} = -\mathbf{A}_{\rm ext}\cdot\hat{n},
\end{equation}
on $\Gamma$, the superconductor boundary.  The scalar potential is decomposed as $\phi = \phi_1 + \phi_2$, satisfying:  
\begin{equation}
\nabla^2\phi_1 = 0, \quad \nabla\phi_1\cdot\hat{n} = -j_{\rm tr}/\sigma \quad \text{on} \ \Gamma,
\end{equation}
and
\begin{equation}
\nabla^2\phi_2 = \frac{1}{\sigma}\nabla\cdot\vec{j}_{\rm sc}, \quad \nabla\phi_2\cdot\hat{n} = 0 \quad \text{on} \ \Gamma.
\end{equation}
The external field is incrementally swept in steps of $B_{\rm z} = 0.51$ mT, with the system relaxed at each step for 700.9 ps or until the convergence criterion is met:
\begin{equation}
\frac{1}{\Omega}\int_\Omega dx \frac{\partial|\psi|^2}{\partial t} < 10^{-9}.
\end{equation}

Figure \ref{fgr:fig1}b presents the free energy difference $\Delta F$ as a function of the applied field. The energy generally increases with increasing field, except at specific values where discrete jumps to lower energy states occur. These jumps correspond to vortex nucleation events.  

Figure \ref{fgr:fig1}c illustrates the order parameter distribution below and above the first energy jump. At $B_{\rm z} = 7.3$ mT, the system remains in a pure superconducting state, exhibiting azimuthal screening currents. At $B_{\rm z} = 12.5$ mT, vortices have nucleated above and below the constriction, forming rows with opposite vorticity and phase windings, behaving as vortices and antivortices. This configuration arises from the structure’s curvature, leading to an inhomogeneous projection of the applied field on the surface normal.  

\begin{figure}[h]%
\centering
\includegraphics[width=0.95\textwidth]{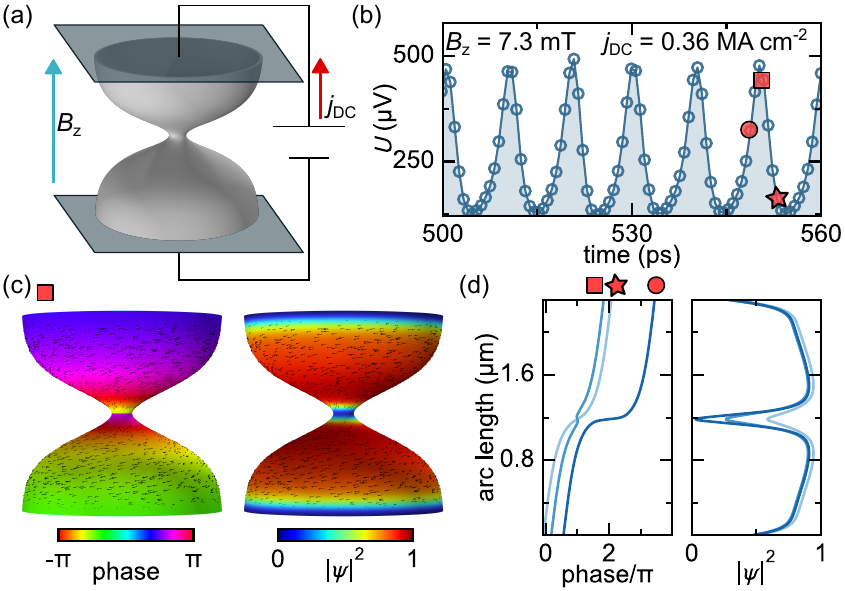}
\caption{Diabolo under applied DC transport current below the lower critical field. (a) Schematic of how the leads are applied. (b) Stable voltage oscillations under DC current. Three points during the oscillations highlighted with red symbols are further analyzed. (c) The spatial distribution of the phase and magnitude of the order parameter during a phase slip at the peak of the voltage. (d) Lineplots of the unwinded phase and order parameter magnitude highlight the suppression of the order parameter and a subsequent jump in phase. 
}
\label{fgr:fig2}
\end{figure}

We next investigate the dynamics under a DC transport current. Current is injected at the edges via appropriate boundary conditions (Figure \ref{fgr:fig2}a), with the applied field set to $B_{\rm z}=7.3$~mT, below the vortex nucleation threshold. Above a critical current, the system enters a dynamic equilibrium characterized by periodic suppression of the order parameter and GHz voltage oscillations, marking the phase slip regime.
Figure \ref{fgr:fig2}b presents the instantaneous voltage $U$ determined as the difference in scalar potential between the top and bottom leads as a function of time at an applied current of $j_{\rm DC}=0.36~\rm{MA~cm^{-2}}$. To understand this regime, we analyze the spatial distribution of the order parameter's phase and magnitude (Figure \ref{fgr:fig2}c). At the voltage peak (indicated by a square symbol), the order parameter is fully suppressed at the constriction. The phase profile is rotationally symmetric and varies along the arc length of the diabolo. Figure \ref{fgr:fig2}d illustrates the phase variation (left) and order parameter (right) along the arc length for three different points in time. The phase undergoes a $2\pi$ jump at the constriction during each phase slip. As described by Michotte \textit{et al.}\cite{michotte_condition_2004}, the periodic phase slip state is stable when the relaxation time of the order parameter phase is shorter than that of the magnitude, i.e., $\tau_\phi<\tau_{|\psi|}$.

\begin{figure}[h]%
\centering
\includegraphics[width=0.95\textwidth]{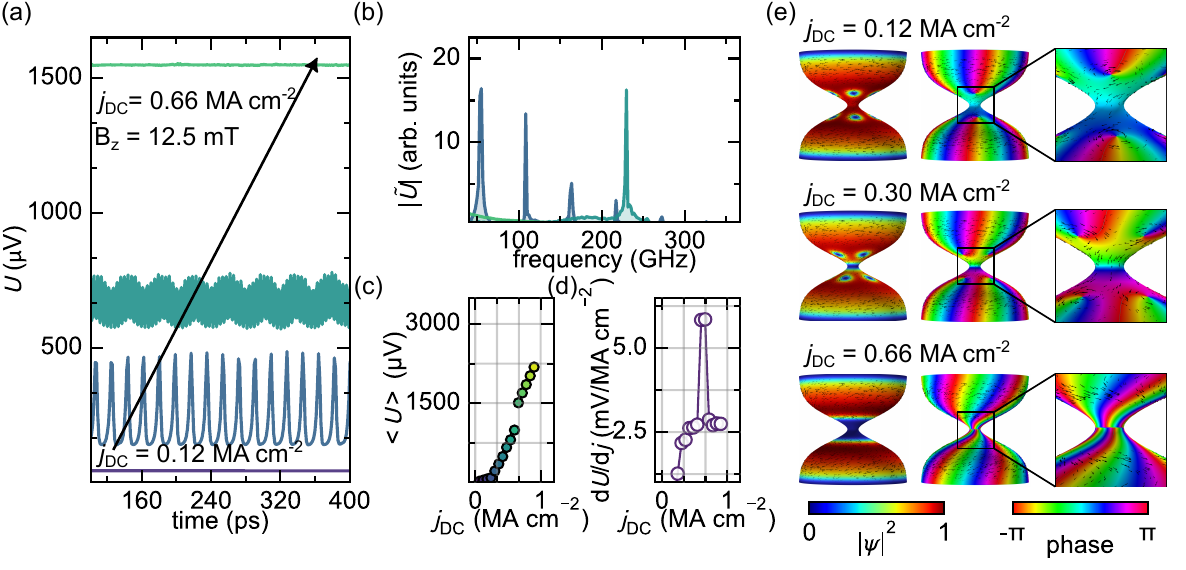}
\caption{DC current sweep in the mixed state. (a) Voltage, time diagram at constant applied field and various applied DC transport currents. (b) The Fourier transformed signals shown in (a). (c) Time averaged voltage as a function of DC current and (d) the differential time averaged voltage as a function of DC current, i.e. differential resistance. (e) Spatial distribution of the order parameter magnitude and phase highlighting the various topologically distinct states at the constriction.}
\label{fgr:fig3}
\end{figure}

The phase slip regime persists in the presence of vortices. Figure \ref{fgr:fig3} presents a DC current sweep performed at an applied field of 12.5~mT, where the diabolo hosts two vortex rows (see Figure \ref{fgr:fig1}c). At low DC currents, vortices and anti-vortices begin to move under the influence of the Lorentz force. Due to the opposite circulation of their screening currents, their motion under the transport current occurs in opposite directions—either clockwise or counter-clockwise. This motion results in a finite, time-constant voltage $U$. As the DC current increases, the diabolo transitions into the phase slip regime. Notably, the frequency of phase slips scales with the DC current (Figure \ref{fgr:fig3}b). This frequency is determined by the Josephson relation, which relates the phase slip frequency $f_{\rm ps}$ to the average voltage drop over the phase slip $\bar{U}=\frac{h}{2e}f_{\rm ps}$\cite{watts-tobin_nonequilibrium_1981}. Additionally, higher-order harmonics appear in the voltage spectrum, which are attributed to the system's intrinsic non-linearity. The time-averaged voltage $\langle U \rangle$ and the corresponding differential voltage $d\langle U \rangle/dj$ are shown in Figure \ref{fgr:fig3}c and \ref{fgr:fig3}d, respectively. Both reveal distinct jumps at $j_{\rm DC}=0.6~\mathrm{MA~cm^{-2}}$ and $j_{\rm DC}=0.66~\mathrm{MA~cm^{-2}}$.  During these transitions, the vortex and anti-vortex rows are absorbed into the constriction by the phase slip. Depending on $j_{\rm DC}$, this process leads to periodic vortex–anti-vortex pair nucleation and annihilation within the constriction (Figure \ref{fgr:fig3}e).

\begin{figure}[h]%
\centering
\includegraphics[width=0.8\textwidth]{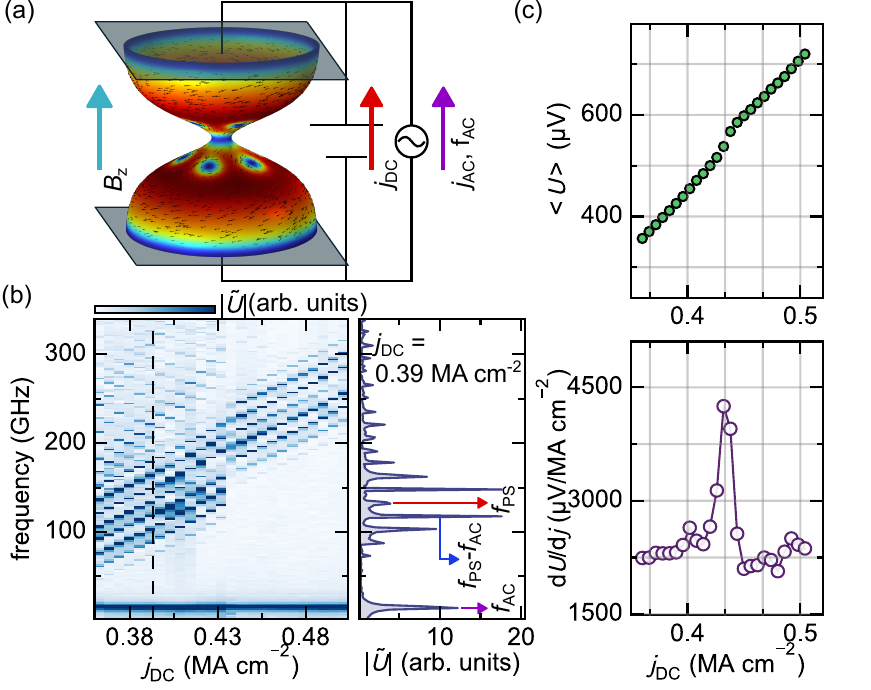}
\caption{DC current sweep of the diabolo under constant applied field and AC current. (a) schematic of the simulation, highlighting that the injected DC transport current is now modulated with an AC current at constant amplitude and frequency. (b) Resulting Fourier transformed voltage $\tilde{U}$ as a function of DC current amplitude. The spectrum shows non-linear frequency mixing of the phase slip frequency $f_{\rm ps}$ with the ac current frequency $f_{AC}$. The right side shows a linecut of the spectrum taken at $j_{DC}=0.39~\mathrm{MA~cm^{-2}}$ (dashed line). (c) Time averaged voltage (top) and differential time averaged voltage (bottom) as a function of DC current.
}
\label{fgr:fig4}
\end{figure}

The phase slips and their DC-current-controlled GHz oscillations open up an intriguing functionality. This is demonstrated by introducing an additional AC current modulation $j_{\mathrm{AC}} = j_0 \sin(2\pi f_{\mathrm{AC}} t)$ to the DC current $j_{\mathrm{DC}}$ when the sample is in the phase slip regime (Figure \ref{fgr:fig4}a). Here, $j_0$ is the AC amplitude, which is fixed at $j_0 = 0.036~\mathrm{MA~cm^{-2}}$, and $f_{\mathrm{AC}} = 15$~GHz is the modulation frequency. We start from an initial state with $B_{\mathrm{z}} = 12.5$~mT and $j_{\mathrm{DC}} = 0.36~\mathrm{MA~cm^{-2}}$, ensuring a stable phase slip regime (Figure \ref{fgr:fig3}). The AC amplitude $j_0$ is gradually ramped from 0 to $0.036~\mathrm{MA~cm^{-2}}$. Subsequently, the AC amplitude and frequency are held constant while sweeping over the DC amplitude. The resulting $j_{\rm DC}$-dependent spectra (Figure \ref{fgr:fig4}b) possess several noteworthy features. Instead of a single fundamental peak and its harmonics, the introduction of the AC signal leads to a range of harmonic peaks. This frequency mixing arises from the nonlinearity of the superconducting condensate, resulting in frequencies of the form $f = n f_{\mathrm{ps}} + m f_{\mathrm{AC}}$, where $n$ and $m$ are integers, and $f_{\mathrm{ps}}$ is the fundamental phase slip frequency driven by the DC current. This mechanism realizes a frequency comb with $n=1$ consisting of hierarchies of equidistant frequency peaks for a fixed $j_{\rm DC}$.  
Additionally, a second set of branches with a steeper slope is observed for small $j_{\rm DC}$ which corresponds to the $n=2$ harmonics. Further simulations indicate that the width of the frequency comb scales with the AC amplitude, while the comb spacing is determined by the modulation frequency $f_{\rm AC}$ (Supporting Information). A distinct jump in the spectrum is observed, which is reflected in both the time-averaged voltage and the differential voltage (Figure \ref{fgr:fig4}c). This discontinuity corresponds to the entrance of a single vortex and anti-vortex pair into the constriction.  

Prior to the jump, the phase evolution resembles that shown in Figure \ref{fgr:fig2}, where the phase distribution within the constriction exhibits rotational symmetry. At the onset of the jump, an isolated vortex first enters the constriction and moves at high velocities. Subsequently, at $t \approx 420$~ps, an anti-vortex enters the constriction. This leads to the disappearance of the higher-order frequency branches in the spectrum. At the higher DC currents shown in Figure \ref{fgr:fig4} b, no additional vortices are observed to enter the constriction. Moreover, upon subsequently reducing the DC current, the system does not exhibit a reappearance of the jump, featuring the presence of hysteresis in the current-voltage characteristic.

The phase slips observed in this study within a multiply connected diabolo structure---characterized by a hollow tube---exhibit distinct behavior compared to those in a 1D wire. At a sufficiently high current, a transition occurs wherein one or more vortices enter the constriction, fundamentally altering the nature of the phase slip. Initially, the phase slip exhibits a quasi-1D character, with a uniform phase along the perimeter. However, after vortex entry, the phase slip is characterized by the periodic annihilation of vortex–antivortex pairs. This transition is marked by a peak in differential resistance and a notable change in the voltage spectrum under AC modulation. Specifically, the absorption of one or more (anti)vortices induces discrete jumps in the spectrum and suppresses higher-frequency modes.

We have investigated the phase slips in the tubular 3D nanostructures by employing the TDGL equations, which are valid for gapless superconductors near the critical temperature \cite{schmid_time_1966}. Previous studies have shown that the onset current for phase slips depends on the relaxation times of both the order parameter's magnitude and phase\cite{vodolazov_dynamic_2002, michotte_condition_2004}. The magnitude relaxation is influenced by the finite inelastic scattering time, which could be incorporated using the generalized TDGL formalism\cite{kramer_theory_1978}. Although this refinement might quantitatively affect the results, the standard TDGL framework has been extensively applied to study phase slips in one- and two-dimensional systems and reliably captured the essential behavior qualitatively \cite{kramer_lossless_1977, andronov_kinematic_1993, baranov_current-voltage_2011}.

\subsection{Conclusion}
We have numerically investigated the static and dynamic properties of three-dimensional Nb diabolo structures, where the constriction radius is comparable to the coherence length ($\xi$). Our primary finding is that a DC-driven diabolo functions as a DC-to-AC converter and, when exposed to an additional AC signal, acts as a frequency comb generator in the microwave regime.

The curved geometry of the diabolo structure supports the formation of both vortices and antivortices in response to a homogeneous magnetic field. At sufficiently high DC currents, a stable phase-slip regime emerges, characterized by the periodic suppression of the order parameter within the constriction. This phase-slip state can coexist with the vortex lattice. As the current increases, the phase-slip region expands and eventually absorbs the vortex rows, resulting in periodic vortex–antivortex annihilation events, which manifest as a distinct peak in the differential voltage. Furthermore, introducing microwave-frequency modulation to the transport current induces frequency mixing and generates sidebands. Notably, when the phase-slip region has absorbed vortices, higher-order frequency modes are suppressed.

Our findings open possibilities for developing 3D integrated, on-chip frequency comb generators by conformal coating of a pillar-like nanotemplate with a single superconducting shell.  The diabolo structure as described here avoids an SIS junction and could also be realized by using a direct write method based on a focused ion beam \cite{cordoba_vertical_2018, porrati_site-selective_2023}. Our design offers a promising advancement for quantum nanotechnologies and high-frequency applications.

\begin{acknowledgement}
The simulations have been performed using the facilities of the Scientiﬁc IT and Application Support Center of EPFL. D.G. acknowledges support by the SNSF via project 227797.

\end{acknowledgement}

\begin{suppinfo}

Additional details on numerical procedure, simulations on wider constrictions, frequency comb dependence on AC and DC current strength.

\end{suppinfo}

\bibliography{bib}

\providecommand{\latin}[1]{#1}
\makeatletter
\providecommand{\doi}
  {\begingroup\let\do\@makeother\dospecials
  \catcode`\{=1 \catcode`\}=2 \doi@aux}
\providecommand{\doi@aux}[1]{\endgroup\texttt{#1}}
\makeatother
\providecommand*\mcitethebibliography{\thebibliography}
\csname @ifundefined\endcsname{endmcitethebibliography}  {\let\endmcitethebibliography\endthebibliography}{}
\begin{mcitethebibliography}{33}
\providecommand*\natexlab[1]{#1}
\providecommand*\mciteSetBstSublistMode[1]{}
\providecommand*\mciteSetBstMaxWidthForm[2]{}
\providecommand*\mciteBstWouldAddEndPuncttrue
  {\def\EndOfBibitem{\unskip.}}
\providecommand*\mciteBstWouldAddEndPunctfalse
  {\let\EndOfBibitem\relax}
\providecommand*\mciteSetBstMidEndSepPunct[3]{}
\providecommand*\mciteSetBstSublistLabelBeginEnd[3]{}
\providecommand*\EndOfBibitem{}
\mciteSetBstSublistMode{f}
\mciteSetBstMaxWidthForm{subitem}{(\alph{mcitesubitemcount})}
\mciteSetBstSublistLabelBeginEnd
  {\mcitemaxwidthsubitemform\space}
  {\relax}
  {\relax}

\bibitem[Wang \latin{et~al.}(2024)Wang, Xu, Li, Shi, Jiang, Guo, Yue, Li, Zhang, Lyu, Pan, Deng, Dong, Tu, Dong, Cao, Zhang, Jia, Sun, Kang, Chen, Wang, Wang, and Wu]{wang_integrated_2024}
Wang,~C.-G. \latin{et~al.}  Integrated and {DC}-powered superconducting microcomb. \emph{Nature Communications} \textbf{2024}, \emph{15}, 4009, Publisher: Nature Publishing Group\relax
\mciteBstWouldAddEndPuncttrue
\mciteSetBstMidEndSepPunct{\mcitedefaultmidpunct}
{\mcitedefaultendpunct}{\mcitedefaultseppunct}\relax
\EndOfBibitem
\bibitem[Likharev(1979)]{likharev_superconducting_1979}
Likharev,~K.~K. Superconducting weak links. \emph{Reviews of Modern Physics} \textbf{1979}, \emph{51}, 101--159\relax
\mciteBstWouldAddEndPuncttrue
\mciteSetBstMidEndSepPunct{\mcitedefaultmidpunct}
{\mcitedefaultendpunct}{\mcitedefaultseppunct}\relax
\EndOfBibitem
\bibitem[Lindelof(1981)]{lindelof_superconducting_1981}
Lindelof,~P.~E. Superconducting microbridges exhibiting {Josephson} properties. \emph{Reports on Progress in Physics} \textbf{1981}, \emph{44}, 949--1026\relax
\mciteBstWouldAddEndPuncttrue
\mciteSetBstMidEndSepPunct{\mcitedefaultmidpunct}
{\mcitedefaultendpunct}{\mcitedefaultseppunct}\relax
\EndOfBibitem
\bibitem[Klapwijk \latin{et~al.}(1977)Klapwijk, Sepers, and Mooij]{klapwijk_regimes_1977}
Klapwijk,~T.~M.; Sepers,~M.; Mooij,~J.~E. Regimes in the behavior of superconducting microbridges. \emph{Journal of Low Temperature Physics} \textbf{1977}, \emph{27}, 801--835\relax
\mciteBstWouldAddEndPuncttrue
\mciteSetBstMidEndSepPunct{\mcitedefaultmidpunct}
{\mcitedefaultendpunct}{\mcitedefaultseppunct}\relax
\EndOfBibitem
\bibitem[Ivlev and Kopnin(1984)Ivlev, and Kopnin]{ivlev_electric_1984}
Ivlev,~B.; Kopnin,~N. Electric currents and resistive states in thin superconductors. \emph{Advances in Physics} \textbf{1984}, \emph{33}, 47--114, Publisher: Taylor \& Francis \_eprint: https://doi.org/10.1080/00018738400101641\relax
\mciteBstWouldAddEndPuncttrue
\mciteSetBstMidEndSepPunct{\mcitedefaultmidpunct}
{\mcitedefaultendpunct}{\mcitedefaultseppunct}\relax
\EndOfBibitem
\bibitem[Kimmel \latin{et~al.}(2017)Kimmel, Glatz, and Aranson]{kimmel_phase_2017}
Kimmel,~G.; Glatz,~A.; Aranson,~I.~S. Phase slips in superconducting weak links. \emph{Physical Review B} \textbf{2017}, \emph{95}, 014518\relax
\mciteBstWouldAddEndPuncttrue
\mciteSetBstMidEndSepPunct{\mcitedefaultmidpunct}
{\mcitedefaultendpunct}{\mcitedefaultseppunct}\relax
\EndOfBibitem
\bibitem[Skocpol \latin{et~al.}(1974)Skocpol, Beasley, and Tinkham]{skocpol_phase-slip_1974}
Skocpol,~W.~J.; Beasley,~M.~R.; Tinkham,~M. Phase-slip centers and nonequilibrium processes in superconducting tin microbridges. \emph{Journal of Low Temperature Physics} \textbf{1974}, \emph{16}, 145--167\relax
\mciteBstWouldAddEndPuncttrue
\mciteSetBstMidEndSepPunct{\mcitedefaultmidpunct}
{\mcitedefaultendpunct}{\mcitedefaultseppunct}\relax
\EndOfBibitem
\bibitem[Andronov \latin{et~al.}(1993)Andronov, Gordion, Kurin, Nefedov, and Shereshevsky]{andronov_kinematic_1993}
Andronov,~A.; Gordion,~I.; Kurin,~V.; Nefedov,~I.; Shereshevsky,~I. Kinematic vortices and phase slip lines in the dynamics of the resistive state of narrow superconductive thin film channels. \emph{Physica C: Superconductivity and its Applications} \textbf{1993}, \emph{213}, 193--199\relax
\mciteBstWouldAddEndPuncttrue
\mciteSetBstMidEndSepPunct{\mcitedefaultmidpunct}
{\mcitedefaultendpunct}{\mcitedefaultseppunct}\relax
\EndOfBibitem
\bibitem[Sivakov \latin{et~al.}(2003)Sivakov, Glukhov, Omelyanchouk, Koval, Müller, and Ustinov]{sivakov_josephson_2003}
Sivakov,~A.~G.; Glukhov,~A.~M.; Omelyanchouk,~A.~N.; Koval,~Y.; Müller,~P.; Ustinov,~A.~V. Josephson {Behavior} of {Phase}-{Slip} {Lines} in {Wide} {Superconducting} {Strips}. \emph{Physical Review Letters} \textbf{2003}, \emph{91}, 267001\relax
\mciteBstWouldAddEndPuncttrue
\mciteSetBstMidEndSepPunct{\mcitedefaultmidpunct}
{\mcitedefaultendpunct}{\mcitedefaultseppunct}\relax
\EndOfBibitem
\bibitem[Mooij and Harmans(2005)Mooij, and Harmans]{mooij_phase-slip_2005}
Mooij,~J.~E.; Harmans,~C. J. P.~M. Phase-slip flux qubits. \emph{New Journal of Physics} \textbf{2005}, \emph{7}, 219--219\relax
\mciteBstWouldAddEndPuncttrue
\mciteSetBstMidEndSepPunct{\mcitedefaultmidpunct}
{\mcitedefaultendpunct}{\mcitedefaultseppunct}\relax
\EndOfBibitem
\bibitem[Liu and Black(2024)Liu, and Black]{liu_performance_2024}
Liu,~M.; Black,~C.~T. Performance analysis of superconductor-constriction-superconductor transmon qubits. \emph{Physical Review A} \textbf{2024}, \emph{110}, 012427\relax
\mciteBstWouldAddEndPuncttrue
\mciteSetBstMidEndSepPunct{\mcitedefaultmidpunct}
{\mcitedefaultendpunct}{\mcitedefaultseppunct}\relax
\EndOfBibitem
\bibitem[Makarov \latin{et~al.}(2022)Makarov, Volkov, Kákay, Pylypovskyi, Budinská, and Dobrovolskiy]{makarov_new_2022}
Makarov,~D.; Volkov,~O.~M.; Kákay,~A.; Pylypovskyi,~O.~V.; Budinská,~B.; Dobrovolskiy,~O.~V. New {Dimension} in {Magnetism} and {Superconductivity}: {3D} and {Curvilinear} {Nanoarchitectures}. \emph{Advanced Materials} \textbf{2022}, \emph{34}, 2101758\relax
\mciteBstWouldAddEndPuncttrue
\mciteSetBstMidEndSepPunct{\mcitedefaultmidpunct}
{\mcitedefaultendpunct}{\mcitedefaultseppunct}\relax
\EndOfBibitem
\bibitem[Fomin and Dobrovolskiy(2022)Fomin, and Dobrovolskiy]{fomin_perspective_2022}
Fomin,~V.~M.; Dobrovolskiy,~O.~V. A {Perspective} on superconductivity in curved {3D} nanoarchitectures. \emph{Applied Physics Letters} \textbf{2022}, \emph{120}, 090501\relax
\mciteBstWouldAddEndPuncttrue
\mciteSetBstMidEndSepPunct{\mcitedefaultmidpunct}
{\mcitedefaultendpunct}{\mcitedefaultseppunct}\relax
\EndOfBibitem
\bibitem[Córdoba and Fomin(2024)Córdoba, and Fomin]{cordoba_topological_2024}
Córdoba,~R.; Fomin,~V.~M. Topological and chiral superconductor nanoarchitectures. \emph{Applied Physics Letters} \textbf{2024}, \emph{124}, 170501\relax
\mciteBstWouldAddEndPuncttrue
\mciteSetBstMidEndSepPunct{\mcitedefaultmidpunct}
{\mcitedefaultendpunct}{\mcitedefaultseppunct}\relax
\EndOfBibitem
\bibitem[Collins \latin{et~al.}(2023)Collins, Rose, and Casaburi]{collins_superconducting_2023}
Collins,~J.~A.; Rose,~C.~S.; Casaburi,~A. Superconducting {Nb} {Nanobridges} for {Reduced} {Footprint} and {Efficient} {Next}-{Generation} {Electronics}. \emph{IEEE Transactions on Applied Superconductivity} \textbf{2023}, \emph{33}, 1--8, Conference Name: IEEE Transactions on Applied Superconductivity\relax
\mciteBstWouldAddEndPuncttrue
\mciteSetBstMidEndSepPunct{\mcitedefaultmidpunct}
{\mcitedefaultendpunct}{\mcitedefaultseppunct}\relax
\EndOfBibitem
\bibitem[Altomare and Chang(2013)Altomare, and Chang]{altomare_onedimensional_2013}
Altomare,~F.; Chang,~A.~M. \emph{One‐{Dimensional} {Superconductivity} in {Nanowires}}, 1st ed.; Wiley, 2013\relax
\mciteBstWouldAddEndPuncttrue
\mciteSetBstMidEndSepPunct{\mcitedefaultmidpunct}
{\mcitedefaultendpunct}{\mcitedefaultseppunct}\relax
\EndOfBibitem
\bibitem[Yerin and Fenchenko(2013)Yerin, and Fenchenko]{yerin_dynamics_2013}
Yerin,~Y.~S.; Fenchenko,~V.~N. Dynamics of the resistive state of a narrow superconducting channel in the ac voltage driven regime. \emph{Low Temperature Physics} \textbf{2013}, \emph{39}, 1023--1031\relax
\mciteBstWouldAddEndPuncttrue
\mciteSetBstMidEndSepPunct{\mcitedefaultmidpunct}
{\mcitedefaultendpunct}{\mcitedefaultseppunct}\relax
\EndOfBibitem
\bibitem[Bogush \latin{et~al.}(2024)Bogush, Dobrovolskiy, and Fomin]{bogush_microwave_2024}
Bogush,~I.; Dobrovolskiy,~O.~V.; Fomin,~V.~M. Microwave generation and vortex jets in superconductor nanotubes. \emph{Physical Review B} \textbf{2024}, \emph{109}, 104516\relax
\mciteBstWouldAddEndPuncttrue
\mciteSetBstMidEndSepPunct{\mcitedefaultmidpunct}
{\mcitedefaultendpunct}{\mcitedefaultseppunct}\relax
\EndOfBibitem
\bibitem[Watts-Tobin \latin{et~al.}(1981)Watts-Tobin, Krähenbühl, and Kramer]{watts-tobin_nonequilibrium_1981}
Watts-Tobin,~R.~J.; Krähenbühl,~Y.; Kramer,~L. Nonequilibrium theory of dirty, current-carrying superconductors: phase-slip oscillators in narrow filaments near {Tc}. \emph{Journal of Low Temperature Physics} \textbf{1981}, \emph{42}, 459--501\relax
\mciteBstWouldAddEndPuncttrue
\mciteSetBstMidEndSepPunct{\mcitedefaultmidpunct}
{\mcitedefaultendpunct}{\mcitedefaultseppunct}\relax
\EndOfBibitem
\bibitem[Rezaev \latin{et~al.}(2020)Rezaev, Smirnova, Schmidt, and Fomin]{rezaev_topological_2020}
Rezaev,~R.~O.; Smirnova,~E.~I.; Schmidt,~O.~G.; Fomin,~V.~M. Topological transitions in superconductor nanomembranes under a strong transport current. \emph{Communications Physics} \textbf{2020}, \emph{3}, 144\relax
\mciteBstWouldAddEndPuncttrue
\mciteSetBstMidEndSepPunct{\mcitedefaultmidpunct}
{\mcitedefaultendpunct}{\mcitedefaultseppunct}\relax
\EndOfBibitem
\bibitem[Du(1994)]{du_finite_1994}
Du,~Q. Finite element methods for the time-dependent {Ginzburg}-{Landau} model of superconductivity. \emph{Computers \& Mathematics with Applications} \textbf{1994}, \emph{27}, 119--133\relax
\mciteBstWouldAddEndPuncttrue
\mciteSetBstMidEndSepPunct{\mcitedefaultmidpunct}
{\mcitedefaultendpunct}{\mcitedefaultseppunct}\relax
\EndOfBibitem
\bibitem[noa()]{noauthor_comsol_nodate}
{COMSOL} {Multiphysics}® v. 6.2. \url{www.comsol.com}\relax
\mciteBstWouldAddEndPuncttrue
\mciteSetBstMidEndSepPunct{\mcitedefaultmidpunct}
{\mcitedefaultendpunct}{\mcitedefaultseppunct}\relax
\EndOfBibitem
\bibitem[Oripov and Anlage(2020)Oripov, and Anlage]{oripov_time-dependent_2020}
Oripov,~B.; Anlage,~S.~M. Time-dependent {Ginzburg}-{Landau} treatment of rf magnetic vortices in superconductors: {Vortex} semiloops in a spatially nonuniform magnetic field. \emph{Physical Review E} \textbf{2020}, \emph{101}, 033306\relax
\mciteBstWouldAddEndPuncttrue
\mciteSetBstMidEndSepPunct{\mcitedefaultmidpunct}
{\mcitedefaultendpunct}{\mcitedefaultseppunct}\relax
\EndOfBibitem
\bibitem[Gao and Xie(2023)Gao, and Xie]{gao_finite_2023}
Gao,~H.; Xie,~W. A {Finite} {Element} {Method} for the {Dynamical} {Ginzburg}–{Landau} {Equations} under {Coulomb} {Gauge}. \emph{Journal of Scientific Computing} \textbf{2023}, \emph{97}, 19\relax
\mciteBstWouldAddEndPuncttrue
\mciteSetBstMidEndSepPunct{\mcitedefaultmidpunct}
{\mcitedefaultendpunct}{\mcitedefaultseppunct}\relax
\EndOfBibitem
\bibitem[Schmid(1966)]{schmid_time_1966}
Schmid,~A. A time dependent {Ginzburg}-{Landau} equation and its application to the problem of resistivity in the mixed state. \emph{Physik der kondensierten Materie} \textbf{1966}, \emph{5}, 302--317\relax
\mciteBstWouldAddEndPuncttrue
\mciteSetBstMidEndSepPunct{\mcitedefaultmidpunct}
{\mcitedefaultendpunct}{\mcitedefaultseppunct}\relax
\EndOfBibitem
\bibitem[Michotte \latin{et~al.}(2004)Michotte, Mátéfi-Tempfli, Piraux, Vodolazov, and Peeters]{michotte_condition_2004}
Michotte,~S.; Mátéfi-Tempfli,~S.; Piraux,~L.; Vodolazov,~D.~Y.; Peeters,~F.~M. Condition for the occurrence of phase slip centers in superconducting nanowires under applied current or voltage. \emph{Physical Review B} \textbf{2004}, \emph{69}, 094512\relax
\mciteBstWouldAddEndPuncttrue
\mciteSetBstMidEndSepPunct{\mcitedefaultmidpunct}
{\mcitedefaultendpunct}{\mcitedefaultseppunct}\relax
\EndOfBibitem
\bibitem[Vodolazov and Peeters(2002)Vodolazov, and Peeters]{vodolazov_dynamic_2002}
Vodolazov,~D.~Y.; Peeters,~F.~M. Dynamic transitions between metastable states in a superconducting ring. \emph{Physical Review B} \textbf{2002}, \emph{66}, 054537\relax
\mciteBstWouldAddEndPuncttrue
\mciteSetBstMidEndSepPunct{\mcitedefaultmidpunct}
{\mcitedefaultendpunct}{\mcitedefaultseppunct}\relax
\EndOfBibitem
\bibitem[Kramer and Watts-Tobin(1978)Kramer, and Watts-Tobin]{kramer_theory_1978}
Kramer,~L.; Watts-Tobin,~R.~J. Theory of {Dissipative} {Current}-{Carrying} {States} in {Superconducting} {Filaments}. \emph{Physical Review Letters} \textbf{1978}, \emph{40}, 1041--1044\relax
\mciteBstWouldAddEndPuncttrue
\mciteSetBstMidEndSepPunct{\mcitedefaultmidpunct}
{\mcitedefaultendpunct}{\mcitedefaultseppunct}\relax
\EndOfBibitem
\bibitem[Kramer and Baratoff(1977)Kramer, and Baratoff]{kramer_lossless_1977}
Kramer,~L.; Baratoff,~A. Lossless and {Dissipative} {Current}-{Carrying} {States} in {Quasi}-{One}-{Dimensional} {Superconductors}. \emph{Physical Review Letters} \textbf{1977}, \emph{38}, 518--521\relax
\mciteBstWouldAddEndPuncttrue
\mciteSetBstMidEndSepPunct{\mcitedefaultmidpunct}
{\mcitedefaultendpunct}{\mcitedefaultseppunct}\relax
\EndOfBibitem
\bibitem[Baranov \latin{et~al.}(2011)Baranov, Balanov, and Kabanov]{baranov_current-voltage_2011}
Baranov,~V.~V.; Balanov,~A.~G.; Kabanov,~V.~V. Current-voltage characteristic of narrow superconducting wires: {Bifurcation} phenomena. \emph{Physical Review B} \textbf{2011}, \emph{84}, 094527\relax
\mciteBstWouldAddEndPuncttrue
\mciteSetBstMidEndSepPunct{\mcitedefaultmidpunct}
{\mcitedefaultendpunct}{\mcitedefaultseppunct}\relax
\EndOfBibitem
\bibitem[Córdoba \latin{et~al.}(2018)Córdoba, Ibarra, Mailly, and De~Teresa]{cordoba_vertical_2018}
Córdoba,~R.; Ibarra,~A.; Mailly,~D.; De~Teresa,~J.~M. Vertical {Growth} of {Superconducting} {Crystalline} {Hollow} {Nanowires} by {He}+ {Focused} {Ion} {Beam} {Induced} {Deposition}. \emph{Nano Letters} \textbf{2018}, \emph{18}, 1379--1386, Publisher: American Chemical Society\relax
\mciteBstWouldAddEndPuncttrue
\mciteSetBstMidEndSepPunct{\mcitedefaultmidpunct}
{\mcitedefaultendpunct}{\mcitedefaultseppunct}\relax
\EndOfBibitem
\bibitem[Porrati \latin{et~al.}(2023)Porrati, Barth, Gazzadi, Frabboni, Volkov, Makarov, and Huth]{porrati_site-selective_2023}
Porrati,~F.; Barth,~S.; Gazzadi,~G.~C.; Frabboni,~S.; Volkov,~O.~M.; Makarov,~D.; Huth,~M. Site-{Selective} {Chemical} {Vapor} {Deposition} on {Direct}-{Write} {3D} {Nanoarchitectures}. \emph{ACS Nano} \textbf{2023}, \emph{17}, 4704--4715, Publisher: American Chemical Society\relax
\mciteBstWouldAddEndPuncttrue
\mciteSetBstMidEndSepPunct{\mcitedefaultmidpunct}
{\mcitedefaultendpunct}{\mcitedefaultseppunct}\relax
\EndOfBibitem
\end{mcitethebibliography}

\end{document}